\documentclass[12pt]{article}
\usepackage{graphicx}
\usepackage{epsfig}
\usepackage{latexsym}
\usepackage{latexsym}
\usepackage{amssymb}
\newcommand{\labell}[1]{\label{#1}}  

\newcommand{\reef}[1]{(\ref{#1})}

\newcommand{\bibbyitem}[1]{\bibitem{#1}}

\DeclareSymbolFont{AMSb}{U}{msb}{m}{n}
\DeclareMathSymbol{\IN}{\mathbin}{AMSb}{"4E}
\DeclareMathSymbol{\IZ}{\mathbin}{AMSb}{"5A}
\DeclareMathSymbol{\IR}{\mathbin}{AMSb}{"52}
\DeclareMathSymbol{\Q}{\mathbin}{AMSb}{"51}
\DeclareMathSymbol{\II}{\mathbin}{AMSb}{"49}
\DeclareMathSymbol{\IC}{\mathbin}{AMSb}{"43}
\DeclareMathSymbol{\IP}{\mathbin}{AMSb}{"50}
\DeclareMathSymbol{\IH}{\mathbin}{AMSb}{"48}
\DeclareMathSymbol\IA{\mathalpha}{AMSb}{"41}
\DeclareMathSymbol\IS{\mathalpha}{AMSb}{"53}

\def\Q{{\cal Q}}

\begin{document}  
  
\bigskip  
\hskip 4.7in\vbox{\baselineskip12pt  
\hbox{hep-th/0107261}}  
  
\bigskip  
\bigskip  
\bigskip  
\bigskip  
\bigskip  
\bigskip 
  
\begin{center}
{\Large \bf The K\"ahler Structure}\\
\bigskip
{\Large \bf of} \\
\bigskip
{\Large \bf Supersymmetric Holographic RG Flows}
  \end{center}
\bigskip  
\bigskip  
\bigskip  
\bigskip 
  
\centerline{\bf Clifford V. Johnson, Kenneth J.
  Lovis, David C. Page}

\bigskip  
\bigskip  
\bigskip

\centerline{\it Centre  
for Particle Theory}
  \centerline{\it Department of Mathematical Sciences}  
\centerline{\it University of  
Durham}
\centerline{\it Durham, DH1 3LE, U.K.}  

\centerline{$\phantom{and}$}  

\bigskip

\centerline{\small \tt  
  c.v.johnson@durham.ac.uk, k.j.lovis@durham.ac.uk, d.c.page@durham.ac.uk}  
  
\bigskip  
\bigskip  
\bigskip  

  
\begin{abstract}  
  \vskip 4pt We study the metrics on the families of moduli spaces
  arising from probing with a brane the ten and eleven dimensional
  supergravity solutions corresponding to renormalisation group flows
  of supersymmetric large $N$ gauge theory.  In comparing the geometry
  to the physics of the dual gauge theory, it is important to identify
  appropriate coordinates, and starting with the case of $SU(N)$ gauge
  theories flowing from ${\cal N}{=}4$ to ${\cal N}{=}1$ {\it via} a
  mass term, we demonstrate that the metric is K\"ahler, and solve for
  the K\"ahler potential everywhere along the flow. We show that the
  asymptotic form of the K\"ahler potential, and hence the peculiar
  conical form of the metric, follows from special properties of the
  gauge theory.  Furthermore, we find the analogous K\"ahler structure
  for the ${\cal N}{=}4$ preserving Coulomb branch flows, and for an
  ${\cal N}{=}2$ flow. In addition, we establish similar properties
  for two eleven dimensional flow geometries recently presented in the
  literature, one of which has a deformation of the conifold as its
  moduli space.  In all of these cases, we notice that the K\"ahler
  potential appears to satisfy a simple universal differential
  equation. We prove that this equation arises for all purely Coulomb
  branch flows dual to both ten and eleven dimensional geometries, and
  conjecture that the equation holds much more generally.
\end{abstract}  
\newpage  
\baselineskip=18pt  
\setcounter{footnote}{0}  
  
  
\section{Introduction}
 
Using a D--brane as a probe of the ten dimensional geometry of stringy
duals\cite{malda, gkp, w1} of gauge theories has proven to be an
interesting and useful avenue of investigation. (There are analogous
statements about the use of M--branes as probes of eleven dimensional
geometries representing M--ish duals of close cousins of gauge
theories.)  There are a number of technical reasons for this, which
all follow from the fact that the low energy action on the D--brane's
world volume is {\it directly} related to a sector of the full gauge
theory at low energy, without any need for employing holographic
dictionaries which are sometimes mysterious or incomplete.  With due
care, this means that the quantities which arise on the brane can
often be interpreted or manipulated with a direct gauge theory
intuition in mind, and then used to shed light on the rest of the ten
dimensional geometry. In short, the probe D--brane's world--volume is
an excellent testbed for finding a good framework, such as more
illuminating coordinates, in which to describe the whole gauge
dual\footnote{There are a number of examples in the literature,
  touching on many of these aspects and issues, for instance
  refs.\cite{bpp,ejp,beh,primer}.}.

In this paper we exhibit further examples of this fact, in the context
of supergravity geometries representing the renormalisation group (RG)
flow\cite{gppz1, dz1} from ${\cal N}=4$ supersymmetric pure large $N$
$SU(N)$ theory in the ultraviolet (UV) to
supersymmetric theories of various sorts, in the infrared (IR).

We will begin with the case of flows which preserve only ${\cal N}=1$
supersymmetry in the dual gauge theory, and revisit a result of
ours\cite{jlp} for the Coulomb branch moduli space metric obtained for
the flow\cite{freed1,pw2} to a large $N$ version of the
Leigh--Strassler point\cite{pilch,robmatt,lsflow}. There were a number of
interesting features of the metric which become most apparent near the
origin of moduli space, where we obtained:
\begin{equation}
ds^2_{{\cal M}_{\rm IR}}={1\over 8\pi^2 g_{\rm YM}^2}\left[{3\over4} du^2+
 u^2\left({4\over 3}\sigma_3^2+
{\sigma_1^2+\sigma_2^2}\right)\right]\ .
\labell{stretched}
\end{equation}
This four dimensional branch represents the allowed vevs of two
complex scalar fields $\phi_1$ and $\phi_2$.  There is a radial
coordinate $u$ and three Euler--type angular coordinates
$(\varphi_1,\varphi_2,\varphi_3)$ introduced {\it via} $\sigma_i$, the
standard left--invariant Maurer--Cartan forms satisfying
$d\sigma_i=\epsilon_{ijk}\sigma_j\wedge\sigma_k$.

The sum $\sum_{i=1}^3\sigma_i^2$ is the standard metric on a round
$S^3$, and so there are stretched $S^3$'s for the radial slices due to
the presence of the $4/3$ instead of unity. While a deformation which
preserves $SU(2)_F\times U(1)_R$ (the global symmetry of the gauge
theory) was to be expected, the gauge theory origin of the precise
number $4/3$, and hence the resulting conical singularity at the
origin, and other features of the metric everywhere along the flow,
all deserve some explanation.

We show that the question is best postponed until one has found better
coordinates for carrying out an investigation of the implicitly
defined quantities of direct relevance to a gauge theory discussion.
The key point is that for a low--energy sigma model, the metric on the
moduli space is the quantity which controls the kinetic terms for the
scalar fields. In superspace, the kinetic terms are written in terms
of a single function, the K\"ahler potential $K$:
\begin{equation}
{\cal L}=\int d^4\theta \, K(\Phi^i,\Phi^{j\dagger})-\left\{\int d^2\theta\,
 W(\Phi^i)
+{\rm h.c.}\right\} \ ,
\end{equation}
where $\Phi^i$ are chiral superfields whose lowest components are the
scalars whose vevs we are exploring, ``h.c.'' means ``hermitian
conjugate', and $W(\Phi)$ is the superpotential. For studying low
energy, it is enough to keep just two derivatives in our effective
action and so this is the form in which we should expect to wish to
write our results\footnote{The D--brane probe also carries a $U(1)$
  gauge field which is decoupled from the scalar fields.  Its gauge
  coupling is given by the ten dimensional dilaton.}.

Our task then turns into one of attempting to prove the existence of a
K\"ahler potential for the probe metric. We succeed in doing this, and
this is a highly non--trivial check on the consistency of the full ten
(and eleven) dimensional flow geometries from the literature which we
study, and of our probe computations. An amusing feature of this
computation is that the flow differential equations themselves
guarantee the existence of the K\"ahler potential. 

Having shown its existence, we can proceed to answer the gauge theory
questions by determining exactly how the potential is constructed out
the basic fields in the theory. This only needs the asymptotic form of
the potential which we can readily write down, and using this, we show
that there is a simple scaling argument which accounts for the precise
form of the metric in the IR.

We can go much further that that, however. We observe that there is a
strikingly simple differential equation which the potential satisfies
along the flow and we are able to solve for it exactly {\it everywhere
  along the flow}!  Moving away from our starting example, we find
similar results for a more general ${\cal N}=1$ preserving flow
presented in the literature recently\cite{warnernew}, finding the same
differential equation for $K$.

Since this relation for $K$ is so simple, we do a survey of many other
ten and eleven dimensional RG flow geometries in the
literature\cite{pw1,bs2,newwarner}, preserving ${\cal N}=2$ and ${\cal
  N}=4$ supersymmetries, and we find that our equation is apparently
universal. We prove directly that it always applies to the case of the
ten and eleven dimensional duals of purely Coulomb--branch
flows\cite{freed2,bakas,bakas2}, and are led to conjecture that it is
universal.  Note that this includes a slight generalisation of the
differential equation which applies to the case of flows from
AdS$_4\times S^7$ and AdS$_7\times S^4$.

Since it is notoriously difficult to obtain exact ${\cal N}=1$ results
for the K\"ahler potential ---such riches are usually reserved for the
superpotential, where holomorphy and non--renormalisation theorems are
powerful allies--- we find our results heartening, since they suggest
that there is much to be gained by applying these methods further in
the quest to extract useful information about strongly coupled gauge
theories from supergravity duals of not inconsiderable complexity.

\section{The Prototype Example: An ${\cal N}=1$  Flow in $D=4$  }
\subsection{The Ten Dimensional Solution}  
\label{tendee}  
The ten dimensional solutions computed in ref.\cite{pw2} describing
the gravity dual of ${\cal N}=4$ supersymmetric $SU(N)$ Yang--Mills
theory, mass deformed to ${\cal N}=1$ in the IR may be written as:
\begin{equation}  
ds^2_{10}=\Omega^2 ds^2_{1,4}+ds^2_{5}\ ,  
\labell{fullmetric}  
\end{equation}  
for the Einstein metric, where  
\begin{equation}  
ds^2_{1,4}=e^{2A(r)}\left(-dt^2+dx_1^2+dx_2^2+dx_3^2\right)+dr^2\ ,  
\labell{littlemetric}  
\end{equation}  
and (see also refs.\cite{cvetic,pilch}):  
\begin{eqnarray}  
ds_5^2&=&{L^2}{\Omega^2\over\rho^2\cosh^2\chi}\left[{d\theta^2}
+\rho^6\cos^2\theta\left({\cosh\chi\over   
    {\bar X}_2}\sigma_3^2+{\sigma_1^2+\sigma_2^2\over 
{\bar X}_1}\right)\right.\nonumber\\  
&&\left.+{{\bar X}_2\cosh\chi\sin^2\theta\over {\bar X}^2_1} \left({d\phi}
+{\rho^{6}\sinh\chi \tanh\chi \cos^2\theta\over {\bar X}_2}
\sigma_3\right)^2\right]\ ,  
\labell{bigmetric}  
\end{eqnarray}  
with  
\begin{eqnarray}  
\Omega^2&=&{{\bar X}_1^{1/2}\cosh\chi\over \rho}\nonumber\\  
{\bar X}_1&=&\cos^2\theta+\rho^6\sin^2\theta\nonumber\\  
{\bar X}_2&=&{\rm sech}\chi\cos^2\theta+\rho^6\cosh\chi\sin^2\theta\ .  
\labell{warp}  
\end{eqnarray}

The functions $\rho(r)\equiv e^{\alpha(r)}$ and $\chi(r)$ 
are the supergravity scalars coupling to certain operators in the dual
gauge theory.  There is a one--parameter family of solutions for them
which gives therefore a family of supergravity solutions. Together
with $A(r)$, they obey the following equations\cite{freed1}:
\begin{eqnarray}  
{d\rho\over dr}&=&{1\over 6L}\rho^2\frac{\partial W}{\partial\rho}={1\over 6L}
\left({\rho^6(\cosh(2\chi)-3)+\cosh(2\chi) + 1\over\rho}\right)  
\nonumber \\  
{d\chi\over dr}&=&{1\over L}\frac{\partial W}{\partial\chi}={1\over 2L}
\left({(\rho^6-2)\sinh(2\chi)\over \rho^2}\right)\nonumber \\  
{dA\over dr}&=&-{2\over 3L} W=-{1\over 6L \rho^2}\left(\cosh(2\chi)(\rho^6-2)
-(3\rho^6+2)\right)\ ,
\labell{flows}  
\end{eqnarray}  
for which no explicit closed form analytic solution is known.
The quantity $W$ is the supergravity superpotential:
\begin{equation}
W=\frac{1}{4}\rho^4(\cosh 2\chi -3) -\frac{1}{2\rho^2}(\cosh 2\chi+1)\ .
\labell{superpotential}
\end{equation}
Note that the asymptotic UV ($r\to+\infty$) behaviour of the fields
$\chi(r)$ and $\alpha(r)=\log(\rho(r))$ is given by\cite{freed1}:
\begin{equation}
\chi(r)\to a_0e^{-r/L}+\ldots\ ;
\qquad \alpha(r)\to {2\over3}a_0^2 {r\over L} e^{-2r/L}+{a_1\over\sqrt{6}}e^{-2r/L}
+\ldots
\labell{asymptone}
\end{equation} 
Notice that this limit gives AdS$_5\times S^5$ in the UV, with
cosmological constant $\Lambda=-6/L^2$ where the normalisations are
such that the gauge theory and string theory quantities are related to
them as:
\begin{eqnarray}
L=\alpha^{\prime1/2} (2g^2_{\rm
  YM}N)^{1/4}\ ;\qquad g^2_{\rm YM}=2\pi g_s\ .
\labell{relations}
\end{eqnarray}
This limit defines the $SO(6)$ symmetric critical point of the ${\cal
  N}=8$ supergravity scalar potential where all of the 42 scalars
vanish which is dual to the ${\cal N}=4$ large $N$ $SU(N)$ gauge
theory.

Meanwhile, in the IR ($r\to-\infty$) the asymptotic behaviour
is\cite{freed1}:
\begin{eqnarray}
&&\chi(r)\to {1\over2}\log 3-b_0e^{\lambda r/L}+\ldots\ ;
\qquad \alpha(r)\to {1\over6}\log 2-{\sqrt{7}-1\over6}b_0 e^{\lambda r/L}
+\ldots\ ,\nonumber \\
&&{\rm where}\quad \lambda={2^{5/3}\over3}(\sqrt{7}-1)\ .
\labell{asympttwo}
\end{eqnarray}
defining another, $SU(2){\times}U(1)$ symmetric, critical point of the
scalar potential\cite{pilch}. It preserves only ${\cal N}=2$
supersymmetry of the maximal ${\cal N}=8$ for five dimensional
supergravity, dual to a conformally invariant ${\cal N}=1$
supersymmetric fixed point $SU(N)$ gauge theory at large $N$. It has
an AdS$_5$ part, but the transverse part of the solution has only
$SU(2){\times}U(1)$, which\footnote{The $SU(2)$ is the
  left--invariance of the $\sigma_i$.  For details of the $U(1)$
  R--symmetry see section \ref{scaling}.}  corresponds to the
$SU(2)_F\times U(1)_R$ symmetry of the theory, which has two adjoint
massless flavours transforming as an $SU(2)_F$ doublet.
  
The fields $\Phi$ and $C_{(0)}$, the ten dimensional dilaton and R--R
scalar, are gathered into a complex scalar field
$\lambda=C_{(0)}+ie^{-\Phi}$ on which $SL(2,{\mathbf Z})$ has a
natural action. This $SL(2,{\mathbf Z})$ is the duality symmetry of
the gauge theory in the UV (the dilaton is related to the gauge
coupling $g_{\rm YM}^{\phantom{2}}$, and the R--R scalar to the
$\Theta$--angle), and an action of it will be inherited by the gauge
theory in the IR. 

The non--zero parts of the two--form potential, $C_{(2)}$, and the
NS--NS two--form potential $B_{(2)}$ are listed in ref.\cite{pw2}, but
we will not need them here.
It was shown in ref.\cite{jlp} that the five form field strength of
the R--R four form potential $C_{(4)}$, presented in ref.\cite{pw2},
to which the D3--brane naturally couples, may be integrated to give a
closed form for the potential, the relevant part of which which we
write as\footnote{By ``relevant'', we mean the part which gets pulled
  back to the D--brane aligned parallel to the $(x^0,x^1,x^2,x^3)$
  directions.}:
\begin{eqnarray} 
C_{(4)}&=& -\frac{4}{g_s}w(r,\theta)\,
 dx_0\wedge dx_1\wedge dx_2 \wedge dx_3\ ,\nonumber \\
{\rm where}\quad  w(r,\theta)&=&
{e^{4A}\over8\rho^2}
[\rho^6\sin^2\theta(\cosh(2\chi)-3)-\cos^2\theta(1+\cosh(2\chi))]\ .
\end{eqnarray}  

\subsection{Moduli Space Metric from a Probe}
  
In ref.\cite{jlp} we worked in static gauge, partitioning the
spacetime coordinates, $x^{\mu}$, according to\footnote{Recall that
  the $\varphi_i$ are angles on the deformed $S^3$ of
  section~\ref{tendee}.}: $x^i=\{x^0,x^1,x^2,x^3\}$, and
$y^m=\{r,\theta,\phi,\varphi_1,\varphi_2,\varphi_3\}$.  We aligned the
brane's worldvolume along the first four spacetime directions and
obtained the effective lagrangian:
\begin{eqnarray}  
{\cal L}\equiv T-V
={\tau_3\over2}
\Omega^2 e^{2A} G_{mn} {\dot y}^m{\dot y}^n
-\tau_3\sin^2\theta e^{4A}\rho^4(\cosh(2\chi)-1)\ ,  
\labell{theresult}
\end{eqnarray}  
where the $G_{mn}$ refer to the Einstein frame metric components, and
we have neglected terms higher than quadratic order in the velocities
in constructing the kinetic term.  The quantity
\begin{equation}
 \tau_3  =  \frac{1}{8\pi^2 g_{YM}^2} \frac{2}{{\alpha'}^2}\ ,
\labell{charge}
\end{equation}
is the D3--brane  charge under the R--R four--form potential.

The moduli space all along the flow is the four dimensional space
$\sin\theta=0$. For $r\to+\infty$ it is simply the flat metric on
$\IR^4$.  In the limit $r\to-\infty$, inserting the IR values of the
functions (see equation (\ref{asympttwo})), using the relations in
equations (\ref{relations}) and defining:
\begin{eqnarray}
u={\rho_0L\over\alpha^\prime}e^{{r/ \ell}}\ ,\qquad \ell={3\over 2^{5/3}}
 L\ ,\qquad
\qquad\rho_0\equiv\rho_{\rm
  IR}=2^{1/6}
\end{eqnarray}
we get equation~\reef{stretched}. 

At this point, as we discussed in the previous section, it is prudent
to stop and see if we can demonstrate the existence of a metric in
K\"ahler form, since this is the low energy metric for an ${\cal N}=1$
supersymmetric sigma model.

\subsection{The Search for K\"ahler Structure}

\label{sec:general1}

The moduli space is parameterised by the vevs of the massless scalars,
which we shall write as $z_1$ and $z_2$.  The $z_i$ transform in the
fundamental of $SU(2)$, while their complex conjugates transform in
the anti--fundamental representation.  The $SU(2)$ flavour symmetry
implies that the K\"ahler potential is a function of $u^2$ only where
we define,
\begin{equation}
  u^2 = z_1 \bar{z}_1 + z_2 \bar{z}_2 \ ,
\end{equation}
and the reader should note that this is not the coordinate $u$ we
defined in the IR in the previous subsection. We shall uncover the
relation between the two shortly.

Dividing the coordinates (and indices) into holomorphic and
anti--holomorphic (those without and those with a bar), if the
K\"ahler structure exists the metric is given by
\begin{eqnarray}
  ds^2 & = & g_{\mu\bar{\nu}} dz^\mu dz^{\bar{\nu}} 
   =  g_{1\bar{1}} dz_1 d\bar{z}_1 + g_{1\bar{2}} dz_1 d\bar{z}_2 + g_{2\bar{1}} dz_2 d\bar{z}_1 + g_{2\bar{2}} dz_2 d\bar{z}_2\ ,
\end{eqnarray}
where
\begin{eqnarray}
  g_{\mu\bar{\nu}} & = & \partial_\mu \partial_{\bar{\nu}} K(u^2) 
   =  \partial_\mu (\partial_{\bar{\nu}} (u^2) K') 
   =  \partial_\mu (\partial_{\bar{\nu}} (u^2)) K' + \partial_\mu (u^2) \partial_{\bar{\nu}} (u^2) K'' \ ,
\end{eqnarray}
where the primes denote differentiation with respect to $u^2$, and we
have inserted our assumption about the $u$ dependence of $K$.  Notice
that since \begin{eqnarray} \partial_i (u^2) & = & \bar{z}_i \quad
  {\rm and} \quad \bar{\partial}_i (u^2) = z_i\ ,
\end{eqnarray}
we have, 
\begin{eqnarray}
  g_{1\bar{1}} & = & \partial_1 \bar{\partial}_1 K 
   =  K' + z_1 \bar{z}_1 K''\ , \nonumber\\
  g_{1\bar{2}} & = & \bar{z}_1 z_2 K''\ ,
\end{eqnarray}
and so on.  So some quick algebra shows that the metric can be written
as
\begin{eqnarray}
  ds^2 & = & (dz_1 d\bar{z}_1 + dz_2 d\bar{z}_2)K' 
   + (\bar{z}_1 dz_1 + \bar{z}_2 dz_2)(z_1 d\bar{z}_1 + z_2 d\bar{z}_2)K''\ .
\end{eqnarray}
Now notice that\footnote{For instance, see p.~377 of ref.~\cite{eguchi}.} 
\begin{eqnarray}
  du & = & \frac{1}{2u} 
(\bar{z}_1 dz_1 + \bar{z}_2 dz_2 + z_1 d\bar{z}_1 + z_2 d\bar{z}_2)\qquad
 {\rm and}\nonumber\\
u \sigma_3 & = & \frac{1}{2u} ( -i \bar{z}_1 dz_1 -i \bar{z}_2 dz_2 + i
 z_1 d\bar{z}_1 + i z_2 d\bar{z}_2)\ ,
\end{eqnarray}
which is convenient, since we can write
\begin{eqnarray}
  du + iu \sigma_3 & = & \frac{1}{u} (\bar{z}_1 dz_1 + \bar{z}_2 dz_2) \quad {\rm and}\quad
  du - iu \sigma_3  =  \frac{1}{u} (z_1 d\bar{z}_1 + z_2 d\bar{z}_2)\ .
\end{eqnarray}
In exchange for a bit more algebra, we arrive at the following form
for our metric:
\begin{equation}
  ds^2 = (K' + u^2 K'')du^2 + u^2(K'(\sigma_1^2 + \sigma_2^2) + (K' + u^2 K'') \sigma_3^2)\ .
  \labell{eq:metric1}
\end{equation}

\subsection{Comparison with probe result}
\label{sec:comparison1}
The result derived using the brane probe should be written out at this
stage, to give:
\begin{equation}
  ds^2 = \frac{\tau_3}{2} \left\{ \frac{\cosh^2 \chi}{\rho^2} e^{2A} dr^2 + L^2 \rho^2 e^{2A}(\cosh^2 \chi \sigma_3^2 + \sigma_1^2 + \sigma_2^2) \right\}\ .
  \labell{eq:probemetric}
\end{equation}
The explicit $SU(2)$ invariance in this equation is that of the
flavour symmetry, so in order to put the metric into K\"ahler form we
need a change of radial coordinate relating $r$ and $u$.

Comparing equations (\ref{eq:metric1}) and (\ref{eq:probemetric}) we
obtain three equations:
\begin{eqnarray}
  (K' + u^2 K'') du^2 & = & \frac{\tau_3}{2} \frac{\cosh^2 \chi}{\rho^2} e^{2A} dr^2\ ,
  \labell{eq:eq1} \\
  u^2 (K' + u^2 K'') & = & \frac{\tau_3}{2} L^2 \rho^2 e^{2A} \cosh^2 \chi\ ,
  \labell{eq:eq2} \\
  u^2 K' & = & \frac{\tau_3}{2} L^2 \rho^2 e^{2A} \ .
  \labell{eq:eq3}
\end{eqnarray}
Using the first two equations we find
\begin{equation}
   dr^2 = \frac{L^2 \rho^4}{u^2} du^2\ .
  \labell{eq:du1}
\end{equation}
A solution is:
\begin{equation}
  u = \frac{L}{\alpha'} e^{f(r)/L}\ ,\quad {\rm with} \quad
\frac{df}{dr}=\frac{1}{\rho^2}\ .
\labell{udef}
\end{equation}
The latter is always positive and so defines a sensible radial coordinate
$u$.  

We can now define K by the differential equation (\ref{eq:eq3}):
\begin{equation}
  K'=\frac{dK}{d(u^2)} = \frac{\tau_3}{2} \frac{L^2 \rho^2 e^{2A}}{u^2}\ ,
  \labell{eq:Kdef}
\end{equation}
and we have to check that such a $K$ obeys equation (\ref{eq:eq2}),
which can be written as:
\begin{equation}
  u^2 \frac{d}{d(u^2)} (u^2 K') = 
\frac{\tau_3}{2} L^2 \rho^2 e^{2A} \cosh^2 \chi\ .
\end{equation}
{}From the definition of $u$ in equation~\reef{udef}, we have that:
\begin{equation}
  \frac{d}{d(u^2)} = \frac{L \rho^2}{2 u^2} \frac{d}{dr}\ ,
  \labell{derivatives}
\end{equation}
and so we need to show
\begin{equation}
  \frac{L \rho^2}{2} \frac{d}{dr} (u^2 K') 
= \frac{\tau_3}{2} L^2 \rho^2 e^{2A} \cosh^2 \chi\ .
  \labell{eq:toprove}
\end{equation}
{}From our definition of $K$ in equation~(\ref{eq:Kdef}) this amounts to
requiring us to show that:
\begin{equation}
  \frac{d}{dr} ( \rho^2 e^{2A}) =  \frac{2}{L }  e^{2A} \cosh^2 \chi\ ,
\end{equation}
which seems like a tall order.  Amazingly, performing the derivative
on the left hand side and substituting the flow equations for
$\rho(r)$ and $A(r)$ listed in \reef{flows} gives {\it precisely} the
result on the right. We have demonstrated the existence of the
K\"ahler potential! In fact, using the equation~\reef{derivatives} we
can write an alternative form for the definition of $K$, to accompany
\reef{eq:Kdef}, which is:
\begin{equation}
  \frac{dK}{dr}={\tau_3}{L}e^{2A(r)}\ .
  \labell{niceK}
\end{equation}

After some thought, one can readily write down an exact solution to
the equation~\reef{niceK} for the K\"ahler potential {\it everywhere
  along the flow}. Up to additive constants, it is:
\begin{equation}
K = { \tau_3 L^2 e^{2A}  \over 4} \left(\rho^2  + {1\over\rho^{4}}\right)\ .
\labell{exactK}
\end{equation}
This is a very simple and satisfying result.

We have found a (family of) K\"ahler manifolds with an $SU(2)\times
U(1)$ holonomy. We can readily compute interesting properties of these
manifolds using the K\"ahler potential. For example, the Ricci form's
components are readily computed to be:
\begin{equation}
R_{{\bar\mu}\nu}=
-2\partial_{\bar\mu}\partial_{\nu}\log\left[K'(K'+u^2K'')\right]\ ,
\end{equation}
which of course generically defines a non--trivial first Chern class.

\subsection{A Few Asymptotic Results}

\subsubsection{Large $u$}
\label{sec:largeu}

For large $u$ ({\it i.e.},~in the limit of large vevs), $\rho \sim 1$
so that, from equation~\reef{udef} we have $u \sim \frac{L}{\alpha'}
\exp(r/L)$, and to leading order:
\begin{equation}
  K \sim \frac{\tau_3}{2} L^2 e^{2r/L} = \frac{1}{8 \pi^2 g_{YM}^2} u^2\ ,
\end{equation}
which implies the expected flat metric:
\begin{equation}
  ds^2 = \frac{1}{8 \pi^2 g_{YM}^2} (du^2 + u^2 (\sigma_1^2 + \sigma_2^2 + \sigma_3^2))\ .
\end{equation}
We can also look at next--to--leading order corrections to the
K\"ahler potential.  Recalling the asymptotic solutions for $\alpha$
and $\chi$ in equations~\reef{asymptone} and also the flow
equations~\reef{flows} one can show:
\begin{equation}
  A(r) \simeq \frac{r}{L} - \frac{a_0^2}{6} e^{-2r/L} + O(e^{-4r/L})\ ,
\end{equation}
so that
\begin{equation}
  K \simeq \tau_3 L^2 \left(\frac{1}{2} e^{2r/L} - \frac{a_0^2}{3} \frac{r}{L} \right) \ ,
\end{equation}
where we have now discarded terms of order $\exp(-2r/L)$ as well as
constant terms.  Similarly, the corresponding expression for $u^2$ is
\begin{equation}
  u^2 \simeq \frac{L^2}{{\alpha'}^2} \left(e^{2r/L} + \frac{4 a_0^2}{3} \frac{r}{L} \right)\ .
\end{equation}
Returning to the K\"ahler potential, we find:
\begin{equation}
  K \simeq \frac{1}{8 \pi^2 g_{YM}^2} \left[u^2 - \frac{a_0^2 L^2}{{\alpha'}^2} \ln\left(\frac{{\alpha'}^2 u^2}{L^2}\right)\right]\ .
\end{equation}

This expression looks similar to that which one might obtain from a
one--loop calculation in field theory. To compare with such a result
we need to know how $a_0^2$ corresponds to the mass for $\Phi_3$.  To
deduce this we can look at the probe result at large $u$ more closely.
The result of the probe calculation was given in
equation~\reef{theresult}.

To leading order, we have 
\begin{eqnarray}
  |z_1|^2 + |z_2|^2 & = & \frac{L^2}{{\alpha'}^2} e^{2r/L} \cos^2 \theta\ ,
\quad\mbox{and}\nonumber\\
  |z_3|^2 & = & \frac{L^2}{{\alpha'}^2} e^{2r/L} \sin^2 \theta\ ,
\end{eqnarray}
and so
\begin{equation}
  {\cal L} = \frac{1}{8\pi^2 g_{YM}^2} \left((|\dot{z}_1|^2 + |\dot{z}_2|^2 + |\dot{z}_3|^2) -  \frac{4 a_0^2}{L^2} |z_3|^2 \right)\ ,
\end{equation}
where the asymptotic solution for $\alpha$ and for $\chi$ have again been used.  The mass of $\Phi_3$  is therefore 
\begin{equation}
  m_3 = \frac{2 a_0}{L}\ .
\end{equation}
Inserting this into the K\"ahler potential, we obtain
\begin{equation}
  K \simeq \frac{1}{8 \pi^2 g_{YM}^2} u^2 
- \frac{N m_3^2}{16 \pi^2} \ln\left(\frac{{\alpha'}^2 u^2}{L^2}\right)\ ,
\end{equation}
which is of the form expected for the tree--level plus  one--loop  correction.

\subsubsection{Small $u$}
\label{infra}

For small $u$, $\rho \to 2^{1/6}$ and we have
\begin{equation}
  u \sim \frac{L}{\alpha'} \exp\left( \frac{r}{2^{1/3} L} \right)\ .
\end{equation}
This gives us the K\"ahler potential:
\begin{equation}
  K \sim \frac{\tau_3}{2} L^2 \frac{3}{2^{5/3}} \left(\frac{u^2 {\alpha'}^2}{L^2}\right)^{4/3} = \frac{1}{8 \pi^2 g_{YM}^2} \frac{3}{2^{5/3}} \left(\frac{{\alpha'}^2}{L^2}\right)^{1/3} (u^2)^{4/3} \ ,
\end{equation}
and so the metric in the IR is:
\begin{equation}
  ds^2 \sim \frac{1}{8 \pi^2 g_{YM}^2} 2^{1/3} \left(\frac{u^2 {\alpha'}^2}{L^2}\right)^{1/3} \left(\frac{4}{3} du^2 + u^2 \left(\sigma_1^2 + \sigma_2^2 + \frac{4}{3} \sigma_3^2 \right)\right) \ ,
\end{equation}
which can be converted to the original form~\reef{stretched} presented
in ref.\cite{jlp} after the redefinition $u\to u^{3/4}$ and an overall
rescaling.

So now we understand that the curious form of the IR metric, noticed
in ref.\cite{jlp}, is simply a consequence of the power, $4/3$, of
$u^2$ which appears in the K\"ahler potential. As we shall see in
section~\ref{scaling}, this power shall follow from a scaling
argument within the field theory. 

\section{A More General ${\cal N}=1$ Flow in $D=4$}

\subsection{The Ten Dimensional Solution}

This time the solution, presented in ref.\cite{warnernew}, allows a
new scalar, $\beta=\log\nu$, to vary.  The flow to non--zero values of
$\beta$ breaks the $SU(2)\times U(1)$ to $U(1)^2$, and $\beta$ in fact
controls a vev on the Coulomb branch of the ${\cal N}=1$ theory which
we explored previously. If instead $\chi$ was to stay zero all along
the flow this is an ${\cal N}=4$ Coulomb branch flow, which we will
examine more closely in section~\ref{coulombic}.
 
The new metric will be of the same broad structure as given in
equation~\reef{fullmetric}, with equation~\reef{littlemetric} for one
component, while the warp factor is given by\cite{warnernew}: 
\begin{equation}
\Omega^2 =   \cosh\chi 
\left( 
 (\nu^2 \cos^2
  \phi + \nu^{-2} \sin^2 \phi) \frac{\cos^2\theta}{\rho^2} + 
\rho^4 \sin^2\theta\right)^{1 \over 2}\ , 
\labell{warped}
\end{equation}
and the  deformed angular metric is given as follows\cite{warnernew}:
\begin{eqnarray}
ds_5^2 = && \frac{L^2}{\Omega^2}
\left[\rho^{-4} \, \big(\cos^2 \theta + \rho^6\, \sin^2 \theta \,
(\nu^{-2}   \cos^2\phi  +\nu^2\, \sin^2\phi) \big)\, d\theta^2 \right.\nonumber\\
&& + \rho^2\cos^2\theta(\nu^2\cos^2\phi+\nu^{-2}\sin^2\phi)d\phi^2  
-2\rho^2 (\nu^2   - \nu^{-2} ) \,\sin\theta\, \cos \theta\,
\sin\phi\cos \phi d\theta d \phi \nonumber\\
&&+\rho^2\cos^2 \theta  (\nu^{-2}\cos^2\phi  d\varphi_1^2  
+ \nu^2 \sin^2\phi  d\varphi_2^2) \left.
+ \rho^{-4} \sin^2 \theta   d\varphi_3^2 \right]\nonumber\\
&& +  \frac{L^2}{\Omega^6} \sinh^2\chi\cosh^2\chi(\cos^2\theta
 (\cos^2\phi  d\varphi_1 - \sin^2\phi  d\varphi_2)  - 
\sin^2\theta  d\varphi_3)^2\ .
\end{eqnarray}
The $U(1)^2$ symmetry is generated by the Killing vectors
$\partial/\partial\varphi_1$ and $\partial/\partial\varphi_2$.
The superpotential for this flow is given by\cite{warnernew}:
\begin{equation}
W = \frac{1}{4}\rho^{4}(\cosh 2\chi  - 3) 
- \frac{1}{4 \rho^{2}}(\nu^{2} + \nu^{-2} ) (\cosh 2\chi  +1) \ ,
\end{equation}
which generalises the superpotential in
equation~\reef{superpotential}.  The equations of motion for the
supergravity fields are:
\begin{eqnarray}
{d\rho\over dr}&=&{1\over 6L}\rho^2\frac{\partial W}{\partial\rho}
={1\over 12L}\left({2\rho^6(\cosh 2\chi-3)
+(\nu^2 + \nu^{-2})(\cosh 2\chi+1)\over\rho}\right)\ ,\nonumber \\
{d\nu\over dr}&=&{1\over 2L}\nu^2\frac{\partial W}{\partial\nu}=
-{1 \over 4L}\left({ (\cosh 2\chi+1)\nu(\nu^2 
- \nu^{-2})\over\rho^2}\right)\ ,\nonumber \\
{d\chi\over dr}&=&{1\over L}\frac{\partial W}{\partial\chi}=
{\sinh 2\chi\over 2L}
 \left({\rho^6 - (\nu^2 + \nu^{-2})\over\rho^2}\right)\ ,\nonumber \\ 
{dA\over dr}&=&-{2\over 3L} W
=-{1\over 6L}\left({\rho^6(\cosh 2\chi-3)
-(\nu^2 + \nu^{-2})(\cosh 2\chi+1)\over\rho^2}\right)\ .
\labell{flows2}
\end{eqnarray}
The authors of ref.\cite{warnernew}  probed the
metric with a D3--brane, with the following result:
\begin{eqnarray}
ds^2 &=& {1 \over 2} \tau_3 e^{2A} 
\left[ \zeta  (  \rho^{-2} \cosh^2\chi  dr^2
+ L^2\, \rho^2  d \phi^2 ) 
+   L^2 \rho^2 (\nu^{-2} \cos^2 \phi 
d\varphi_1^2 
+ \nu^2 \sin^2 \phi  d\varphi_2^2)\right.\nonumber \\ 
& & \left. \qquad\qquad\qquad\qquad\qquad+  L^2 \rho^2  
\sinh^2\chi \zeta^{-1} (  \cos^2\phi \, d\varphi_1  - 
 \sin^2\phi  \,d\varphi_2 )^2  \right]  \ , \nonumber\\
\mbox{where}\,\,\,\, \zeta &\equiv&(\nu^2 \cos^2 \phi + 
\nu^{-2} \sin^2 \phi )\ .
 \labell{newmoduli}
\end{eqnarray}

\subsection{A K\"ahler Potential}

Again we wish to use ${\cal N}=1$ supersymmetry and the flavour and
R--symmetries of the theory to choose a special set of coordinates in
which the action for the brane probe can be compared to field theory
expectations. In this case the $SU(2){\times}U(1)$ symmetry has been
broken to $U(1)^2$ by giving a vev to one of the massless fields. 

The $U(1)^2$ symmetries are given by constant shifts in $\varphi_1$
and $\varphi_2$. We wish to find a complex structure in which this
metric is K\"ahler and the $U(1)^2$ symmetries are realised linearly.
Therefore, we should choose something like
\begin{equation}
z_{1} = \sqrt{u(r,\phi)} e^{i\varphi_1}\ ,
\qquad z_{2} = \sqrt{v(r,\phi)} e^{-i\varphi_2}\ ,
\end{equation}
and K\"ahler potential 
\begin{equation}
K = K(z_1 {\bar z}_1 , z_2 {\bar z}_2)  = K(u,v)  \ .
\end{equation}
Proceeding as before we write down the form of the metric which
results from this K\"ahler potential. A short calculation gives
\begin{eqnarray}
ds^2 &=& {1 \over 4u^2}\left(u {\partial \over \partial u}\right)^2 K du^2 
+ {1 \over 2uv}\left( u {\partial \over \partial u}\right)
\left(v {\partial \over \partial v}\right) K  du dv 
+ {1 \over 4v^2}\left(v {\partial \over \partial v}\right)^2 K dv^2 \nonumber\\
&+& \left(u {\partial \over \partial u}\right)^2 K d\varphi_1^2 
-2 \left(u {\partial \over \partial u}\right)
\left(v {\partial \over \partial v}\right) K d\varphi_1 d\varphi_2 
+ \left(v {\partial \over \partial v}\right)^2 K d\varphi_2^2\ .
\end{eqnarray} 
Comparison with equation \reef{newmoduli} gives the following set of
equations for $u,v$ and $K$.
\begin{eqnarray}
\left(u {\partial \over \partial u}\right)^2 K &=& {\tau_3 \over 2} e^{2A} L^2 \rho^2 (\nu^{-2} \cos^2 \phi  + \sinh^2 \chi  \zeta^{-1} \cos^4 \phi ) \nonumber\\
&=& 4u^2 e^{2A} \zeta \left(\rho^{-2} \cosh^2 \chi  \left({\partial r \over \partial u}\right)^2 + L^2 \rho^2  \left({\partial \phi \over \partial u}\right)^2 \right)\ , \nonumber\\
\left(v {\partial \over \partial v}\right)^2 K &=& {\tau_3 \over 2} e^{2A} L^2 \rho^2 (\nu^{2} \sin^2 \phi  + \sinh^2 \chi  \zeta^{-1} \sin^4 \phi ) \nonumber\\
&=& 4v^2 e^{2A} \zeta \left(\rho^{-2} \cosh^2 \chi  \left({\partial r \over \partial v}\right)^2 + L^2 \rho^2  \left({\partial \phi \over \partial v}\right)^2 \right)\ , \nonumber\\
\left(u {\partial \over \partial u}\right)\left(v {\partial \over \partial v}\right) K &=& {\tau_3 \over 2} e^{2A} L^2 \rho^2  \sinh^2 \chi  \zeta^{-1} \cos^2 \phi  \sin^2 \phi \nonumber\\
&=& 4uv e^{2A} \zeta \left(\rho^{-2} \cosh^2 \chi  \left({\partial r \over \partial u}\right) \left({\partial r \over \partial v}\right) + L^2 \rho^2  \left({\partial \phi \over \partial u}\right) \left({\partial \phi \over \partial v}\right)\right)\ .
\end{eqnarray}
The solutions for $u$ and $v$ are 
\begin{equation}
u = f(r) \cos^2 \phi\ , \qquad v = g(r) \sin^2 \phi\ ,  \nonumber
\end{equation}
where
\begin{equation}
{df \over dr} = {2 \nu^2 \over {L \rho^2}}f\ , \qquad {dg \over dr} = {2 \over {L \rho^2 \nu^2}}g\ .
\labell{N1-coulomb-vars}
\end{equation} 
We find an exact solution for the K\"ahler potential:
\begin{equation}
K = {\tau_3 \over 2} L^2 e^{2A} \left( \rho^2(\nu^2 - \nu^{-2}) \sin^2 \phi  + {1 \over 2} (\rho^2 \nu^{-2} + \rho^{-4}) \right)\ + a\log(u) + b\log(v) + d\ ,
\end{equation}
where $a,b$ and $d$ are constants.  As before the equations of
motion~\reef{flows2} were needed in order to find a solution. The
specific combinations of the equations used are rather simple and we
reproduce them below:
\begin{eqnarray} 
{d (e^{2A} \rho^2 \nu^{-2}) \over d r} & = & {d (e^{2A} \rho^2 \nu^{2}) \over d r} =  {2 \over L} e^{2A} \cosh^2 \chi \ , \labell{N1-coulomb-a} \nonumber\\
{d (e^{2A} \rho^{-4}) \over dr} & = & {e^{2A} \over L} (3 - \cosh(2\chi))\ .
\labell{N1-coulomb-b}
\end{eqnarray}
In fact, this allows us to write down a remarkably simple solution for
$A$ as a function of $\rho$ and~$\nu$ for the case $\nu\neq1$:
\begin{equation}
e^{2A} =  {k \over \rho^2 (\nu^2 - \nu^{-2})}\ ,
\labell{N1-coulomb-c}
\end{equation}
where $k$ is a constant.  Using this expression and setting $a=b=d=0$
we can simplify the K\"ahler potential and we find again that it
satisfies the equation~\reef{niceK}
\begin{equation}
  K =  {\tau_3 \over 2} L^2 k \sin^2 \phi  + {\tau_3 L^2 e^{2A} \over 4} 
\left(\frac{\rho^2}{\nu^2} + \frac{1}{\rho^4}\right)\ .
\labell{N1-coulomb-K}
\end{equation}

\section{Pure Coulomb Branch Flows}
\label{coulombic}

\subsection{Switching off the Mass in $D=4$}

In fact, it is quite interesting to study a special case of the above.
Let us switch off the mass deformation by setting $\chi = 0$
everywhere. This means that we are studying a purely ${\cal N} = 4$
Coulomb branch deformation.  If we set $\chi = 0$ then equations
\reef{N1-coulomb-a} and \reef{N1-coulomb-b} simplify to yield the
result
\begin{equation}
  e^{2A} \rho^2 \nu^2 =  e^{2A} \rho^2 \nu^{-2} + k = e^{2A} \rho^{-4} + l\ ,
\end{equation}
where $k$ is the constant appearing in equation \reef{N1-coulomb-c}
and $l$ is another integration constant.  We can also write down a
solution for equation \reef{N1-coulomb-vars} in this case:
\begin{equation}
  f =  {L^2 \over {\alpha'}^2} e^{2A} \rho^2 \nu^{-2}, \qquad  g = {L^2 \over {\alpha'}^2} e^{2A}  \rho^2 \nu^2\  .
\end{equation}
Now if we substitute into the expression \reef{N1-coulomb-K} for the
K\"ahler potential, we find
\begin{eqnarray}
  K &=& {\tau_3 \over 2} L^2 e^{2A} ( \rho^2 \nu^2 \sin^2\phi + \rho^2 \nu^{-2} \cos^2\phi ) \nonumber \\
    &=& \frac{1}{8 \pi^2 g_{YM}^2} (u + v) \nonumber \\
    &=& \frac{1}{8 \pi^2 g_{YM}^2} (z_1 \bar{z}_1 + z_2 \bar{z}_2)\ .
\labell{flatspace}
\end{eqnarray}
This is the expected probe result\cite{primer} for a two complex
dimensional subspace of the flat three complex dimensional moduli
space which exists for the ${\cal N} = 4$ theory on Coulomb branch.
Notice that in the standard supergravity flow coordinates, this result
is not manifest, but there is a change of coordinates (achievable with
the aid of the flow equations\footnote{See also
  refs.\cite{beh,freed2,bakas}.})  to make it so.  The ten dimensional
flow metric in the new coordinates is then nothing else but a
distributional D3--brane solution of ref.\cite{kraus}, for which the
transverse space is flat space multiplied by an harmonic function.

Let us see if we can write the general result for the pure Coulomb branch
flows, knowing that the underlying structure of the ten dimensional
solution is so simple.

\subsection{The General Case of the Coulomb Branch}

The general $\cal{N} =$ 4 supersymmetric Coulomb branch flows are
discussed in refs.\cite{freed2,bakas}. The five dimensional
supergravity equations describing these flows were given in
ref.\cite{freed2} and a way to decouple these equations in order to
find exact solutions was presented in ref.\cite{bakas}.  The
ten--dimensional metric corresponds to a continuous distribution of
D3--branes\cite{kraus} and so must take the form
\begin{equation}
  ds^2 = H^{-{1 \over 2}} 
\eta_{\mu \nu} dx^\mu dx^\nu + H^{1 \over 2} (dy_1 ^2 + dy_2 ^2 + \ldots + 
dy_6 ^2)\ ,
\end{equation}
for some harmonic function $H(y_i)$. A D3--brane probing this
background has a flat metric on moduli space\cite{primer}:
\begin{eqnarray}
ds^2 &=& {\tau_3 \over 2} (dy_1 ^2 + dy_2 ^2 + \ldots + dy_6 ^2) 
  = {\tau_3 \over 2} (dz_1 d\bar{z}_1 + dz_2 d\bar{z}_2 + dz_3 d\bar{z}_3)\ ,
\end{eqnarray}
coming from a K\"ahler potential
\begin{equation}
  K = {\tau_3 \over 2} (y_1 ^2 + y_2 ^2 + \ldots + y_6 ^2)\ .
\end{equation}
In ref.\cite{bakas} it was shown that it is natural to write the lift
solution in terms of a radial coordinate~$F$ which satisfies, in the 
coordinates of our discussion:
\begin{equation}
  {dF \over dr} = 2L e^{2A}\ ,
  \labell{Feqn}
\end{equation}
and is related to the $y_i$ by
\begin{equation}
  y_i = (F - b_i)^{1 \over 2} \hat{x_i}\ ,
\end{equation}
where $b_i$ are constants and $\hat{x_i}$ are coordinates on a unit
$S^5$. Changing to $F$ coordinates we find that the K\"ahler potential
is given by
\begin{equation}
K = {\tau_3 \over 2} \sum_i (F-b_i) \hat{x_i}^2 = {\tau_3 \over 2}(F - \sum_i b_i \hat{x_i}^2)\ ,
\end{equation}
from which we can read 
\begin{equation}
  {\partial K \over \partial F} = {\tau_3 \over 2} \ , 
\end{equation}
which when combined with equation~(\ref{Feqn}) results in our
equation~(\ref{niceK}) once again.  In fact it is straightforward to
generalise these results to the case of M2-- and M5--branes on their
analogues of the Coulomb branch by adapting the results of
ref.\cite{bakas2}.  In these cases we find
\begin{equation}
 { \partial K \over \partial r} = \tau_{M2} L e^A\ ,
\labell{LPtwo}
\end{equation}
for M2--branes and
\begin{equation}
 { \partial K \over \partial r} = \tau_{M5} L e^{4A}\ ,
\labell{LPthree} 
\end{equation}
for M5--branes. We conjecture that together with the
equation~\reef{niceK} for systems built of D3--branes, the natural
K\"ahler potentials governing the metric on moduli space for all
supersymmetric holographic RG flows coming from gauged supergravities,
satisfy these equations. We have proven them for purely Coulomb branch
flows, shown them for the Leigh--Strassler flow and a generalisation,
and in the next two sections demonstrate that it is also true for
another three families of examples.

\section{An ${\cal N}=2$ Flow in $D=4$}

We now briefly revisit the ten--dimensional ${\cal N} = 2$
supersymmetric flow solution constructed in ref.\cite{pw1}, which has
been studied {\it via} brane probing in refs.\cite{bpp,ejp,beh}. Again
we wish to find an explicit form of the K\"ahler potential and check
that equation~(\ref{niceK}) is satisfied.  In this case the moduli
space for a D3--brane probe is a one complex dimensional space and so
it is simple to find coordinates in which the metric is K\"ahler,
since these are just the coordinates in which the metric is
conformally flat:
\begin{equation}
 ds^2 = \partial \bar{\partial} K \, dz \, d\bar{z}\ .
\end{equation}
In fact such coordinates have already been found in
refs.\cite{bpp,ejp}, and we reproduce the result here
\begin{equation}
  ds^2 = {\tau_3 \over 2} k^2 L^2 {c \over (c+1)^2} \, dz \, d\bar{z}\ ,
\end{equation}
where
\begin{equation}
c = \cosh(2\chi)\quad \mbox{ and  }\quad 
 z = e^{-i \phi} \sqrt{{(c+1) \over (c-1)}}\ .
\end{equation}
Note that this fixes the complex structure on moduli space but that in
this case the flavour symmetries are insufficient to pin down a unique
choice of complex coordinates. $\cal{N} =$ 2 supersymmetry is used in
ref.\cite{bpp} to match the scalar kinetic term with the kinetic term
of the $U(1)$ gauge field on the brane and so fix a unique set of
coordinates\footnote{See also ref.\cite{beh} for further work and
  generalisations.}.  However, for our purposes we only need the
correct complex structure and so $\cal{N} = $ 1 arguments are enough.
To proceed we set $u = z \bar{z}$ and find
\begin{eqnarray}
 ds^2 = {d \over du} \left( u {d \over du} \right) K \, dz \, d\bar{z} 
      = {\tau_3 \over 2} k^2 L^2 {(u^2 - 1) \over 4 u^2} \, dz \, d\bar{z}\ ,
\end{eqnarray}
which has solution 
\begin{equation}
 K = {\tau_3 \over 2} 
k^2 L^2 \left({1 \over 4} (u - u^{-1}) + a \log(u) + b \right)\ ,
\end{equation}
where $a$ and $b$ are constants. To get this expression for the K\"ahler
potential into the form we want, we need to use the following
solutions\cite{pw1} for the supergravity fields $A$ and $\rho$,
\begin{eqnarray}
  e^A &=& k { \rho^2 \over \sinh2\chi } \nonumber\\
  \rho^6 &=& \cosh 2\chi + \sinh^22\chi 
\left( \gamma + \log \left[{ \sinh\chi \over \cosh \chi } \right] \right)\ .
\end{eqnarray}
If we choose $a = -{1 \over 2}$ and $b = \gamma$ then the K\"ahler
potential simplifies to
\begin{equation}
  K = {\tau_3 \over 2} L^2 \rho^2 e^{2A}\ ,
\end{equation}
which on applying the relevant supergravity equations of motion 
\begin{eqnarray}
\rho\frac{d\rho}{d r}&=&
\frac{1}{3L}\left( \frac{1}{\rho^2}-\rho^4\cosh 2\chi\right)\ , \quad 
\frac{d\chi}{dr}= -\frac{1}{2L}\rho^4\sinh 4\chi\ ,\nonumber\\
\frac{d A}{d r}&=& \frac{2}{3L}\left( \frac{1}{\rho^2}+
\frac{1}{2}\rho^4\cosh 2\chi\right)\ ,
\end{eqnarray}
can be shown to satisfy
equation~\reef{niceK}.

\section{Two ${\cal N}=2$ Flows in $D=3$}
\label{M2sec}

Finally, we examine two non--trivial examples in eleven dimensions
which appeared in the literature very recently\cite{newwarner}. The
first is a flow from the $SO(8)$ invariant 2+1 dimensional fixed point
in the UV (dual to AdS$_4\times S^7$) to an $SU(3)\times U(1)$ fixed
point generalising the fixed point $D=4$ field theory of section~2. It
is dual to an ${\cal N}=2$ gauged supergravity critical
point\cite{nicolai}, and aspects of the dual field theory have been
considered in refs.\cite{ahn}. The second is closely related (from the
eleven dimensional point of view), being an alternative lift of the
{\it same} four dimensional ${\cal N}=8$ supergravity fields as the
first example to a different eleven dimensional completion for which
the moduli space in the limit of large fields is the conifold instead
of $\IR^8$. This is achieved by replacing the stretched $S^5$'s (which
generalise the stretched $S^3$'s in sections~2 and~3) with a space
which is topologically $T^{1,1}$; the authors of ref.\cite{newwarner}
noticed that the same gauged supergravity equations of motion result.
We make a few more comments about the field theory dual of this second
example towards the end of subsection~\ref{connie}.

\subsection{The $S^5$ Flow}
\label{lslike}
The flow under consideration has a 3 (complex) dimensional moduli
space with $SU(3)$ symmetry and we start by finding the general form
for an $SU(n)$ invariant K\"ahler metric on $\mathbb{C}^n$. Let $w^1,
w^2, \ldots, w^n$ be coordinates on $\mathbb{C}^n$. $SU(n)$ invariance
implies that the K\"ahler potential $K$ depends only on the combination
\begin{equation}
  q=w^1\bar{w}^1 + w^2\bar{w}^2 + \ldots + w^n\bar{w}^n\ .
\end{equation}
Let us reparametrise $\mathbb{C}^n \sim \mathbb{R}^{2n} \sim
\mathbb{R}^+ \times S^{2n-1}$ with coordinates $\hat{x}^1, \ldots
,\hat{x}^{2n} ,q$ . The $\hat{x}$'s are coordinates on an $S^{2n-1}$
of unit radius and are related to the $w$'s by
\begin{equation}
w^J = \sqrt{q} (\hat{x}^{2J-1} + i \hat{x}^{2J})\ . 
\end{equation}
A short calculation gives the K\"ahler metric in these coordinates as:
\begin{equation}
ds^2 = {1 \over 4q^2} \left(q {d \over dq}\right)^2 K\, dq^2 + \left(q {d \over dq}\right) K\, d\hat{x}^I d\hat{x}^I + \left(q^2 {d^2 \over dq^2}\right) K \, (\hat{x}^I J_{IJ} d\hat{x}^J)^2\ ,
\labell{SU(n)metric}
\end{equation}
where $J$ is an antisymmetric matrix with $J_{12} = J_{34} = \ldots =
J_{2n-1 \, 2n} = 1$ .

Now we wish to compare this to the metric on moduli space for the flow
solution in ref.\cite{newwarner}. The lift ansatz for the eleven
dimensional metric in ref.\cite{newwarner} is:
\begin{equation}
ds^2_{11} = \Delta^{-1} (dr^2 + e^{2A(r)} ( \eta_{\mu \nu} dx^\mu dx^\nu)) 
+ \Delta^{{1 \over 2}} L^2 ds^2(\rho, \chi) \ ,
\labell{11dmetric}
\end{equation}
where $ds^2(\rho , \chi)$ is given in equation (4.10) of
ref.\cite{newwarner} and we will give the restriction of it to moduli
space below. From this form of the eleven dimensional metric and a
short calculation to determine the location of the moduli space we can
read off the metric on moduli space:
\begin{equation}
ds^2 = {\tau_{M2} \over 2} e^A (\Delta^{-{3 \over 2}} dr^2 
+ L^2 ds^2(\rho , \chi)\arrowvert_{\rm moduli})\ ,
\labell{moduli}
\end{equation}
where 
\begin{equation}
 \Delta^{-{3 \over 2}} = {\cosh^2 \chi \over \rho^2}  \ ,
\end{equation}
on moduli space and:
\begin{equation}
 ds^2(\rho , \chi)\arrowvert_{\rm moduli} = \rho^2 d\hat{x}^I d\hat{x}^I 
+ \rho^2 \sinh^2 \chi (\hat{x}^I J_{IJ} d\hat{x}^J)^2\ , \quad I,J=1\ldots 3\ ,
\labell{moduli2}
\end{equation}
which defines a family of stretched\footnote{Note that $d\hat{x}^I
  d\hat{x}^I$ is the metric on a round $S^5$ while $(\hat{x}^I J_{IJ}
  d\hat{x}^J)^2$ is the $U(1)$ fibre in the description of $S^5$ as a
  $U(1)$ fibre over $\IC\IP^2$. So the function $ \sinh^2 \chi$
  controls a stretching of the fibre in a precise generalisation of
  the case for the $S^3$'s in equation~\reef{eq:probemetric}.} $S^5$'s
generalising the stretched $S^3$'s of our $D=4$ gauge theory result in
equation~\reef{eq:probemetric}.

Now we can find the complex structure and K\"ahler potential which
produce such a metric on moduli space.  Substituting
equation~(\ref{moduli2}) into~(\ref{moduli}) and comparing with
equation~(\ref{SU(n)metric}), we get three equations for the two
unknown functions $q(r)$ and $K(r)$:
\begin{eqnarray}
 {1 \over 4q^2} \left(q {d \over dq}\right)^2 K  dq^2 &=& {\tau_{M2} \over 2}
 e^A \Delta^{-{3 \over 2}} dr^2\ ,
  \labell{M2eqn-a} \\
 \left(q {d \over dq}\right) K &=& {\tau_{M2} \over 2} e^A L^2 \rho^2\ ,
  \labell{M2eqn-b} \\
 \left(q^2 {d^2 \over dq^2}\right) 
K &=& {\tau_{M2} \over 2} e^A L^2 \rho^2 \sinh^2 \chi\ .
 \labell{M2eqn-c}
\end{eqnarray}
Consistency of these equations requires that
\begin{equation}
{d \over dr} (e^A \rho^2) = {2 \over L} e^A \cosh^2 \chi\ ,
\end{equation}
which is a consequence of the supergravity equations of motion. Then
the solutions for $K$ and~$q$ are:
\begin{equation}
K = {\tau_{M2} L^2 \over 4} e^A \left(\rho^{2}+\frac{1}{\rho^6}  \right)\ , \qquad
{dq \over dr} = {2 \over L \rho^2} q\ .
  \labell{KM2}
\end{equation}
We observe again that, happily, this K\"ahler potential satisfies
equation~\reef{LPtwo}, and the contrast with equation~\reef{exactK}
for the analogous four dimensional flow should be noted.

\subsection{The $T^{1,1}$ flow}
\label{connie}
Interestingly, there is a second lift of the same four dimensional
gauged supergravity solution which led to the eleven dimensional
example of the previous section. This was also constructed in
ref.~\cite{newwarner}. The second eleven dimensional geometry is
similar to the first, but with the stretched five spheres ($U(1)$
bundles over $\IC\IP^2$) replaced by stretched $T^{1,1}$'s, ($U(1)$
bundles over $S^2\times S^2$). The details are in
ref.~\cite{newwarner} and since we shall only be interested in the
moduli space of an M2--brane probe, we shall not go into details here.

The metric on moduli space for an M2--brane probing this geometry is
again given by equation~\reef{moduli} except that
equation~\reef{moduli2} is replaced by
\begin{equation}
 ds^2(\rho , \chi)
\arrowvert_{\rm moduli} = \rho^2 ds^2 _{T^{1,1}} 
+ \rho^2 \sinh^2 \chi {1 \over 9} (d\psi + \cos \theta_1 d \phi_1 
+\cos \theta_2 d \phi_2)^2 \ ,
 \labell{moduli3}
\end{equation}
and the metric on $T^{1,1}$ is~\cite{candelas}:
\begin{equation}
 ds^2 _{T^{1,1}} = {1 \over 9} (d\psi + \cos \theta_1 d \phi_1 
+\cos \theta_2 d \phi_2)^2 + {1 \over 6} (d \theta_1 ^2 
+ \sin ^2 \theta_1 d \phi_1 ^2) +{1 \over 6} (d \theta_2 ^2 
+ \sin ^2 \theta_2 d \phi_2 ^2)\ .
 \labell{T11metric}
\end{equation}
In particular, the  $S^5$'s in equation~\reef{moduli2} have
been replaced by $T^{1,1}$'s  so that for
large~$r$ ($\rho \to 1$~, $\chi \to 0$) the two moduli spaces approach
flat $\mathbb{R}^6$, and the Ricci flat K\"ahler conifold,
respectively.

To proceed we first note that the $SU(2) \times SU(2)$ symmetry of the
conifold metric is preserved for all values of $r$. Therefore, we need
to find the general form of an $SU(2) \times SU(2)$ invariant K\"ahler
metric on the conifold and compare to equations~\reef{moduli},
\reef{moduli3}. Such metrics have been studied in ref.~\cite{candelas}
and we shall rederive a result of that paper below.

The conifold is a surface in $\mathbb{C}^4$ parametrised by four
complex coordinates $z_1, \, z_2, \, z_3$ and $z_4$, which satisfy an
equation
\begin{equation}
  z_1 ^2 + z_2 ^2 + z_3 ^2 + z_4 ^2 = 0.
\labell{conifoldeqn}
\end{equation}
An $SU(2) \times SU(2) = SO(4)$ invariant metric depends only on the
combination
\begin{equation}
  p = z_1 \bar{z}_1 + z_2 \bar{z}_2 + z_3 \bar{z}_3 + z_4 \bar{z}_4 .
\end{equation}
Thus, to construct an $SU(2) \times SU(2)$ invariant metric on the
conifold we can start with our equation~\reef{SU(n)metric} for an
$SU(4)$ invariant metric on $\mathbb{C}^4$ and then restrict to the
conifold using equation \reef{conifoldeqn}. For convenience and in
order to adjust a couple of notations we reproduce the equation for an
$SU(4)$ invariant metric on $\mathbb{C}^4$ here:
\begin{equation}
ds^2 = {1 \over 4p^2} \left(p {d \over dp}\right)^2 K\, dp^2 + \left(p {d \over dp}\right) K\, d\hat{z_i} d\hat{\bar{z_i}} + \left(p^2 {d^2 \over dp^2}\right) K \, | \hat{z_i} d\hat{\bar{z_i}}|^2  ,
\end{equation}
where the $\hat{z_i}$'s parametrise an $S^7$, {\it i.e.}~$\Sigma \hat{z_i}
\hat{\bar{z_i}} = 1$. In fact it will be more convenient to work in
terms of a radial coordinate $q = {3 \over 2} p^{2/3}$ for reasons
which will become clear. In these coordinates the $SU(4)$ invariant
metric becomes:
\begin{equation}
ds^2 = {1 \over 4q^2} \left(q {d \over dq}\right)^2 K\, dq^2 + \left(q {d \over dq}\right) K\, \left[{2 \over 3} d\hat{z_i} d\hat{\bar{z_i}} - {2 \over 9} | \hat{z_i} d\hat{\bar{z_i}}|^2 \right] + {4 \over 9} \left(q^2 {d^2 \over dq^2}\right) K \,  | \hat{z_i} d\hat{\bar{z_i}}|^2  .
\end{equation}
Finally, we need to restrict to the conifold by applying equation
\reef{conifoldeqn} to the $\hat{z_i}$'s. If we reparametrise in terms
of the coordinates\footnote{The reader should refer to
  ref.~\cite{candelas} for the explicit form of these coordinates in
  terms of the $\hat{z}_i$'s.}  on $T^{1,1}$ introduced in
equation~\reef{T11metric}, then we find the following form for the
general $SU(2) \times SU(2)$ invariant metric on the conifold:
\begin{equation}
ds^2 = {1 \over 4q^2} \left(q {d \over dq}\right)^2 K\, dq^2 + \left(q {d \over dq}\right) K\, [ds^2 _{T^{1,1}}] + {1 \over 9} \left(q^2 {d^2 \over dq^2}\right) K \, (d\psi + \cos \theta_1 d \phi_1 +\cos \theta_2 d \phi_2)^2  .
\end{equation}
It is now straightforward to compare this metric with the metric on
moduli space~\reef{moduli}, \reef{moduli3}, to extract equations for
$K$ and $q$. The equations which $K$ and $q$ satisfy are precisely
\reef{M2eqn-a}, \reef{M2eqn-b} and \reef{M2eqn-c} with
solution~\reef{KM2}. It should be noted, however, that $q$ has a very
different definition than it did in the flow of the previous section.

\section{Scaling Dimensions and R--symmetries}
\label{scaling}

The supergravity flows discussed in sections 2 and 6 interpolate
between geometries of the form conformal to AdS$\times M$,
corresponding to flows between conformal field theories. We have
already fixed unique coordinates on moduli space by demanding that the
supersymmetry and flavour symmetries be realised linearly in the brane
probe action. Now we should check whether the scaling dimensions of
the chiral superfields are correctly reproduced in these coordinates.

\subsection{Ten dimensional ${\cal N} = 1$ Flow}
\label{scaling-10d}

A simple example to start with is the UV end of the flow in section 2
which is just the standard $AdS_5 \times S^5$ geometry. The $AdS_5$
part of the metric is
\begin{equation}
  ds^2 = e^{2A} \eta_{\mu \nu} dx^\mu dx^\nu + dr^2 \qquad \mbox{where  } A = {r \over L}\ ,
\end{equation}
which has a scaling symmetry under 
\begin{equation}
  x \rightarrow {1 \over \alpha} x \qquad e^{A} \rightarrow \alpha e^{A}\ .
  \labell{scale}
\end{equation}
In fact this is a symmetry of all the fields in the supergravity
solution and thus a symmetry of the action of a brane probing this
background.  In terms of the coordinates on moduli space derived in
section 2, we have $u \sim e^A$ for large $r$ and so the scaling
symmetry becomes
\begin{equation}
 x \rightarrow {1 \over \alpha} x \qquad u \rightarrow \alpha u\  .
\end{equation}
In other words the fields on moduli space have scaling dimension 1
which matches with the field theory prediction for the scalar
components of these chiral superfields in the ${\cal N}=4$ theory.
Next we consider the IR end of the flow solution. Here the solution
again has the scaling symmetry~(\ref{scale}) except that $A = {2^{5/3}
  r \over L}$ in this case. The coordinate $u$ goes like $u \sim \exp
\left( {r \over 2^{1/3} L} \right) \sim (e^A)^{3/4}$ and thus the
scaling symmetry becomes
\begin{equation}
 x \rightarrow {1 \over \alpha} x \qquad u \rightarrow \alpha^{3/4} u\ .
\end{equation}
Therefore, we see that the massless fields have scaling dimension 3/4
here.  Again this agrees with the field theory. To see this we briefly
reproduce the analysis of ref.~\cite{robmatt, freed1,lsflow}.  The ${\cal
  N}=4$ theory in the UV has superpotential
\begin{equation}
  W = h\, \mbox{Tr} ([\Phi_1 , \Phi_2] \Phi_3)\ ,
\end{equation}
and the flow under consideration corresponds to adding a mass term
\begin{equation}
  \delta W = {m \over 2} \mbox{Tr} \Phi_3 ^2 .
\end{equation}
The $\beta$--functions for $h$ and $m$ are given by
\begin{eqnarray}
  \beta_h &=& h (d_1 + d_2 + d_3 - 3) \nonumber\\
  \beta_{m} &=& m (2 d_3 - 3)
\end{eqnarray}
where $d_i$ is the scaling dimension of the scalar component of
$\Phi_i$.  The vanishing of the $\beta$--functions (along with the $SU(2)$
symmetry) requires $d_3 = {3 \over 2}$, $d_1 = d_2 = {3 \over 4}$.

A further check on the coordinates is to consider the $U(1)$
R--symmetry which forms a part of the superconformal group. The
R--charge of the scalar component of $\Phi_i$ is given by $r_i = {2
  \over 3} d_i$ and so $r_1 = r_2 = {1 \over 2}$, $r_3 = 1$ . This
should match the extra $U(1)$ symmetry of the supergravity solution in
ref.~\cite{pw2}. To identify this $U(1)$ symmetry we need to consider
the antisymmetric tensor field (eqn.  3.19 in ref.~\cite{pw2})
\begin{equation}
C_{(2)} = e^{-i \phi} (a_1 d\theta - a_2 \sigma_3 - a_3 d\phi) \wedge (\sigma_1 - i \sigma_2) .
\end{equation}
This has a $U(1)$ symmetry under 
\begin{equation}
  \phi  \rightarrow \phi + \gamma  \qquad \sigma_1 - i \sigma_2 \rightarrow e^{i \gamma} ( \sigma_1 - i \sigma_2) . 
\end{equation}
The charges of the scalar fields $z_1$, $z_2$ under this symmetry
indeed reproduce the field theory values.\footnote{It is interesting
  to consider the space $z_1 = z_2 = 0$ parametrised by $z_3$, the vev
  of the massive scalar. This corresponds to the two--dimensional
  space with $\cos\theta = 0$.  $z_3$ has R--charge $r_3 = 1$ and is
  an $SU(2)$ scalar and therefore is of the form $z_3 = \sqrt{q}
  e^{-i\phi}$. We might guess that a natural choice for the radial
  coordinate q is to be found by putting a D--brane probe at $z_1 = z_2
  = 0$ and considering its kinetic term for motions in the $z_3$
  direction.  If we put this probe metric into the form $ds^2 = g \,
  dz_3 \, d\bar{z}_3$, for some function $g$, we find that
  ${dq / dr} = {2 \rho^4 q / L}.$
%
At the UV fixed point $r \to \infty$, $z_3$ has scaling dimension 1.
For the IR $r \to -\infty$, one finds that $z_3$ has dimension 3/2,
which matches the value deduced above.}

Working the other way, consider the K\"ahler potential at either end
(UV or IR) of the flow.  From the $SU(2)$ flavour symmetry we know
that $K$ is a function of $u^2$ only.  We also know the scaling
dimension of $u^2$.  The K\"ahler term in the action is of the form:
\begin{equation}
  S = \int d^D\! x \,\, \partial_\varphi \, \partial_{\bar{\varphi}} K \, \partial_\mu \varphi \, \partial^\mu \bar{\varphi}
  \labell{Kaction}
\end{equation}
where $\varphi$ are the massless scalars with some scaling dimension
and $D=4$.  For $S$ to be invariant under the scaling symmetry,
classically $K(u^2)$ must have scaling dimension 2.  At the UV end of
the flow $u$ has scaling dimension 1, so $K \sim u^2$, as expected.
At the IR end of the flow solution, $u$ has scaling dimension 3/4 and
so $K \sim (u^2)^{4/3}$.  This matches the result found in section
\ref{infra}.  It is therefore possible to recover the form of the
K\"ahler potential at either end of the flow from a classical scaling
argument.

\subsection{Eleven dimensional ${\cal N} = 2$ Flows}
\label{scaling-11d}

Now we repeat the analysis for the eleven dimension flow solution in
the coordinates of section~6. First let us consider the field theory
which in the UV contains four chiral superfields $\phi_i$, with $ i =
1\ldots4$. The scaling dimensions ($d_i$) of the fields satisfy
\begin{equation}
  d_1 + d_2 + d_3 + d_4 = 2\ ,
\end{equation}
and so $d_i = {1 \over 2}$ by symmetry.  The flow solution corresponds
to perturbing the superpotential by a mass term for $\Phi_4$
\begin{equation}
 \delta W = {m \over 2} \mbox{Tr} \Phi_4 ^2\ .
\end{equation}
This leads to a $\beta$--function 
\begin{equation}
 \beta_{m} = m (2 d_4 - 2)\ . 
\end{equation}
Thus the IR values of the scaling dimensions are $d_4 = 1$,
$d_1=d_2=d_3={1 \over 3}$. Again we have a $U(1)$ R--symmetry forming
part of the superconformal group and in this case the charges satisfy
$r_i = d_i$.

Let us compare these results with the symmetries of the limits of the
flow solution written in the coordinates of section 6. The UV end of
the flow is just $AdS_4 \times S^7$ and has a symmetry under the
scaling \reef{scale}, with $A = {2r \over L}$. The radial coordinate
on moduli space from section 6 is $\sqrt{q} \sim \exp({r \over L})
\sim (e^A)^{1/2}$ and so the scaling symmetry becomes
\begin{equation}
 x \rightarrow {1 \over \alpha} x \qquad \sqrt{q} \rightarrow \alpha^{{1/2}} \sqrt{q} .
\end{equation}
At the IR end of the flow $A \sim {3^{{3/4}} r \over L}$ and $\sqrt{q}
\sim \exp({r \over 3^{{1/4}} L}) \sim (e^A)^{1/3}$. The scaling
symmetry is
\begin{equation}
 x \rightarrow {1 \over \alpha} x \qquad \sqrt{q} \rightarrow \alpha^{{1/3}} \sqrt{q} .
\end{equation}
These are all in agreement with the scaling dimensions of the massless
scalar fields derived above.

Finally we wish to match the extra $U(1)$ symmetry of the supergravity
solution to the R--symmetry of the field theory. To identify the
action of the $U(1)$ symmetry we look at the equation after (4.28) in
ref.\cite{newwarner},
\begin{equation}
 A^{(3)} = {1 \over 4} \sinh{\chi} 
e^{i(-4 \psi -3 \phi)} 
(e^5 - i e^{10}) \wedge (e^6 - i e^9) \wedge (e^7 - i e^8)\ , 
\end{equation}
where the $e^a$ define a frame which locally diagonalises the metric,
and are listed in ref.~\cite{newwarner}. The three form potential
$A^{(3)}$ has a $U(1)$ symmetry with
\begin{equation}
  \psi \rightarrow \psi - 
\gamma \qquad \phi \rightarrow \phi + {4 \over 3} \gamma
\end{equation} 
under which the coordinates $w_i$ have charge ${1 \over 3}$ in
agreement with the field theory.\footnote{Again it interesting to
  consider the two--dimensional space parametrised by the vev of the
  massive scalar; this time it is $w_4$.  One finds that $w_4 =
  \sqrt{t} e^{-i\psi}$ and so to put the probe metric into the form
  $ds^2 = g \, dw_3 \, d\bar{w}_3$ we require
  ${dt / dr} = {2 \rho^6 t / L}$. 
It is then easy to show that the scaling dimensions at the UV or IR
fixed points match those found above.}

We can now consider the K\"ahler potential at either end of the flow,
as we did at the end of section \ref{scaling-10d}. In this case
$SU(3)$ symmetry implies that $K$ is a function of $q$ only.  Again,
we know the scaling dimension of $\sqrt{q}$.  Referring to equation
\reef{Kaction}, but now with $n=3$, one can see that classically $K$
should have scaling dimension 1.  At the UV end of the flow, since
$\sqrt{q}$ has scaling dimension 1/2, $K \sim q$ as before.  For the
IR end, $\sqrt{q}$ has scaling dimension 1/3, and so $K$ should obey
$K \sim q^{3/2}$.  Let us compare this with the results we found in
section~\ref{M2sec},  equation~\reef{KM2}.  We see that
for $r \to -\infty$, $K \sim e^A$ and $q \sim (e^A)^{2/3}$.  This
implies $K \sim q^{3/2}$, matching the result from the classical
scaling argument above. Note, however, that the scaling argument does
not give the K\"ahler potential for arbitrary $r$ (where the scaling
symmetry no longer holds).

For completeness, we note that it is possible to extract scaling
dimensions for the UV and IR ends of the conifold flow of section
\ref{connie}, which provides data about the dual field theories. Since
the large field limit of the flow gives the moduli space as the
conifold, it is natural to assume that the dual field theory will be
an orbifold, with the M2--branes at the origin. The conifold arises as
the moduli space of vacua of a set of fields restricted by a D--term
equation which is the defining equation~\reef{conifoldeqn} of the
conifold. Then, a good description of the field theory on moduli space
is in terms of fields $A_i$ and $B_j$, as in ref.\cite{kw}
(called $X_i$ and $Y_j$ in ref.~\cite{mp}). There is an additional
complex field, say $\Phi$, which has a superpotential giving it a
mass, which drives the flow.  Carrying out the supergravity analysis
as above, we find that the scaling dimensions of the $A$'s and $B$'s
are 3/8 in the UV and 1/4 in the IR.  These numbers are simply 3/4
times the dimensions of the fields in the other flow. A naive guess
for the superpotential which couples all the fields is $W \sim
\epsilon^{ij}\epsilon^{kl}{\rm Tr}(A_i B_k A_j B_l \Phi)$, for which
the $\beta$--function vanishes if $\Phi$ has dimension $1/2$ in the
UV, and 1 in the IR, the latter also fitting nicely with a mass term
$m{\rm Tr}\Phi^2$.

\section{Closing Remarks}

In a previous paper\cite{jlp} we studied the moduli space of a
D--brane probe of a particular ten dimensional geometry dual to a flow
from the ${\cal N}=4$ large $N$ $SU(N)$ gauge theory to an ${\cal
  N}=1$ fixed point. We found a remarkably simple form for the metric
on moduli space. Asymptotically, its peculiar conical form was of
particular interest, since the value of the numerical coefficients
entering the metric demanded a field theory explanation.

In this paper we set out to find suitable coordinates in which to
exhibit the dual supergravity geometries in order to address such
questions, and we found many interesting structures.  We are able to
put the metric on moduli space for a brane probe into a manifestly
K\"ahler form, which is quite natural in  studies of
supersymmetric gauge theory. One amusing point in these calculations
has been the way in which the first order supergravity
equations\footnote{These are derived from Killing spinor equations
  (see e.g. ref.~\cite{freed1}) or a Bogomol'nyi--type argument (see
  ref.~\cite{bakas2} and references therein).} which ensure the ${\cal
  N} = 1$ supersymmetry of the full supergravity geometry are also
precisely what is needed to make the metric on moduli space K\"ahler.

Furthermore, we have shown that by working in these natural coordinates
we can give a simple derivation of the scaling dimensions of the
massless chiral superfields. The usual methods for calculating
dimensions of these fields from the dual supergravity involve
linearising fluctuations about a given background and is
computationally rather tedious. The most direct benefit of this
derivation for the purposes of this paper was the ease with which it
allows us to acheive our initial goal: The K\"ahler metric controls
all of the coefficients in the asymptotic metric, and their precise
values follow from the scaling argument.

We studied many other cases, preserving various amounts of
supersymmetry, and in both ten and eleven dimensions. In every case
that we studied, it was possible to find an {\it exact} expression for
the K\"ahler potential in terms of the supergravity fields. On the
one hand, this is very exciting, since for a low amount of
supersymmetry, exact results for this quantity are harder to find than
for {\it e.g.,} the superpotential, where holomorphicity is a powerful
constraint.  On the other hand, we note that in translating to the
field theory we find that the exact supergravity expression for the
K\"ahler potential does not always tell the whole story, or in the
most simple way, and so we advise some caution on the part of the
user.  For instance, in the example of the maximal supersymmetry
preserving Coulomb branch flows, the K\"ahler potential has a very
simple form from the field theory viewpoint.  However, it is looks
more complicated in terms of the supergravity scalars. One has to
unpack the details of the supergravity fields and/or change variables
to another set of coordinates in which the simplicity is more
manifest\cite{bakas,bakas2,beh,kraus}, to see that in those cases
there is in fact a simpler story to be told.  Another example comes
from the two eleven dimensional flow geometries of section
\ref{M2sec}.  One simple expression for the supergravity K\"ahler
potential describes two different theories, the details of which must
still be unpacked by using the knowledge of the detailed behaviour of
the scalars.

In each of the examples we have considered, the K\"ahler potential
satisfies a remarkable simple differential equation \reef{niceK} (or
\reef{LPtwo}, \reef{LPthree} in eleven dimensions), which we
conjecture to be universal for these sorts of holographic RG flows (we
proved them to be so for all the Coulomb branch flows).  It will be
interesting to try to understand better the origins of this equation
in the gauged supergravity and also its direct interpretation in the
dual gauge theory.

\section*{Acknowledgements} 
We thank Nick Evans, David Tong and Nick Warner for comments. We also
thank Nick Warner for showing us a prelimiary version of
ref.\cite{newwarner}. KJL and DCP would like to thank Douglas Smith
for extensive discussions, and Ian Davies, James Gregory and Antonio
Padilla for other helpful discussions.  CVJ would like to thank Columbia
University's Physiology and Physics Departments for kind occasional
use of a printer, and SJB for her patience.
KJL and DCP are funded by the Engineering and Physical Sciences
Research Council.  This paper is report number DCTP--01/61 at the CPT,
Durham.


\end{document}